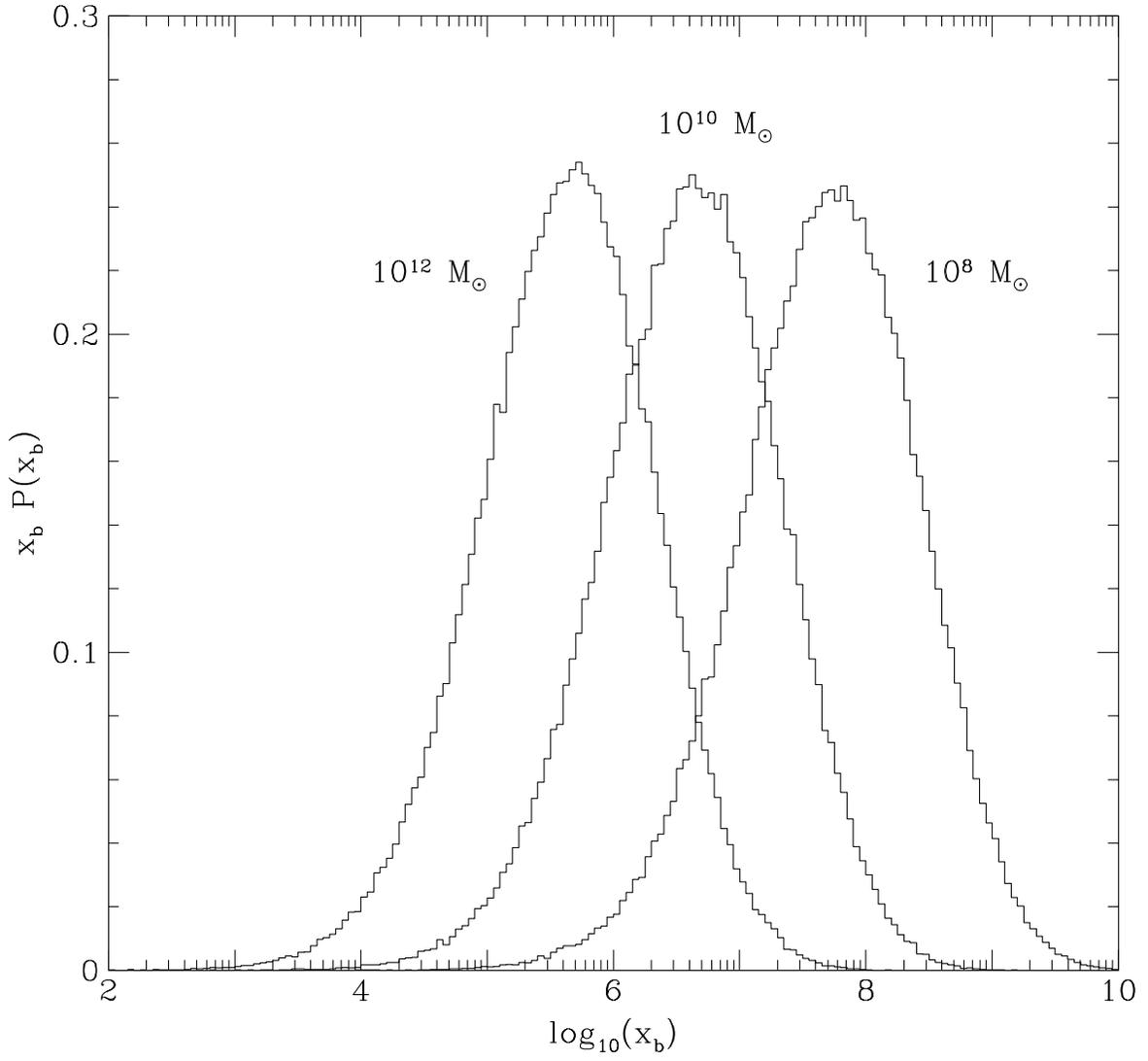

Fig. 1

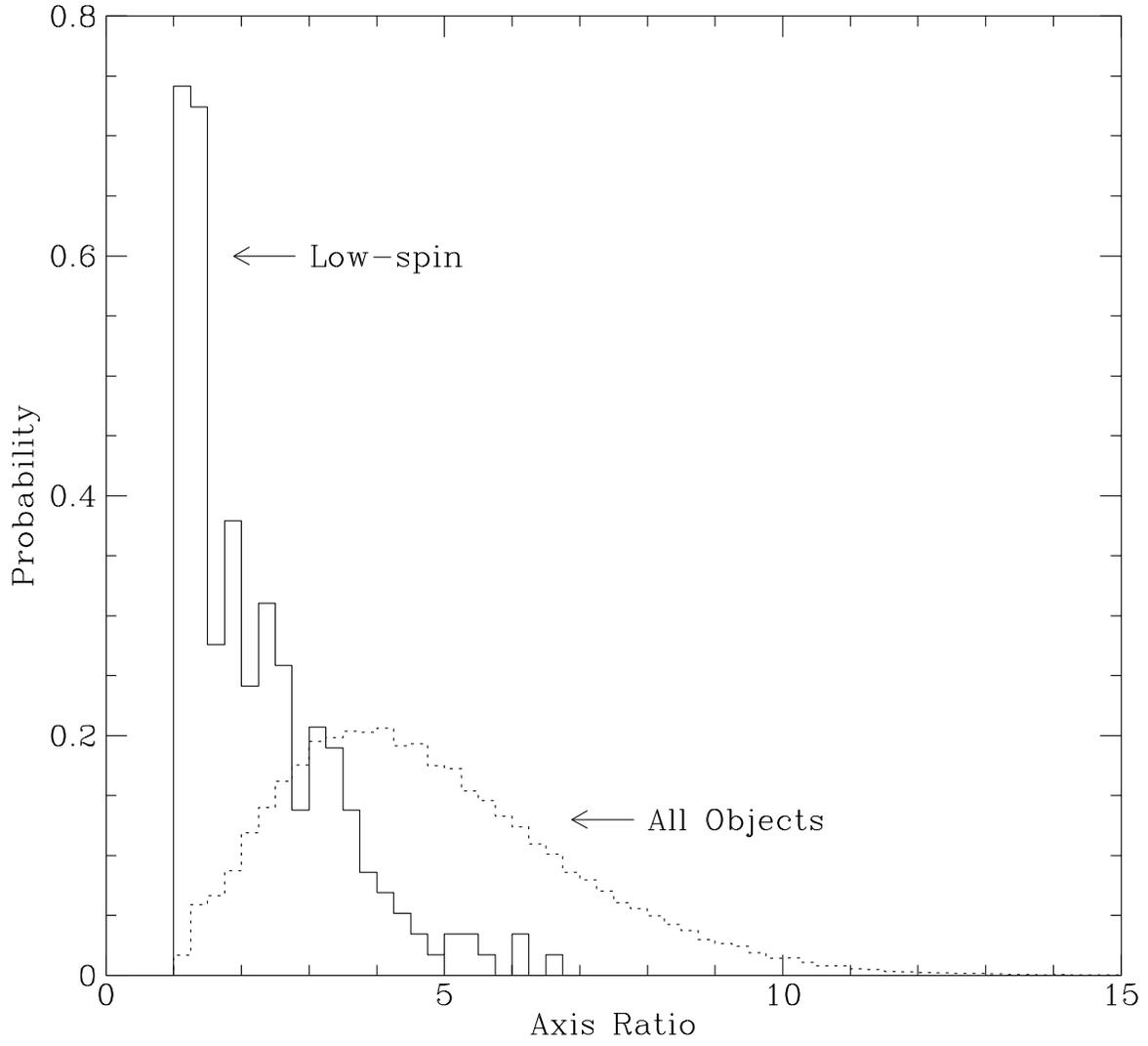

Fig. 2

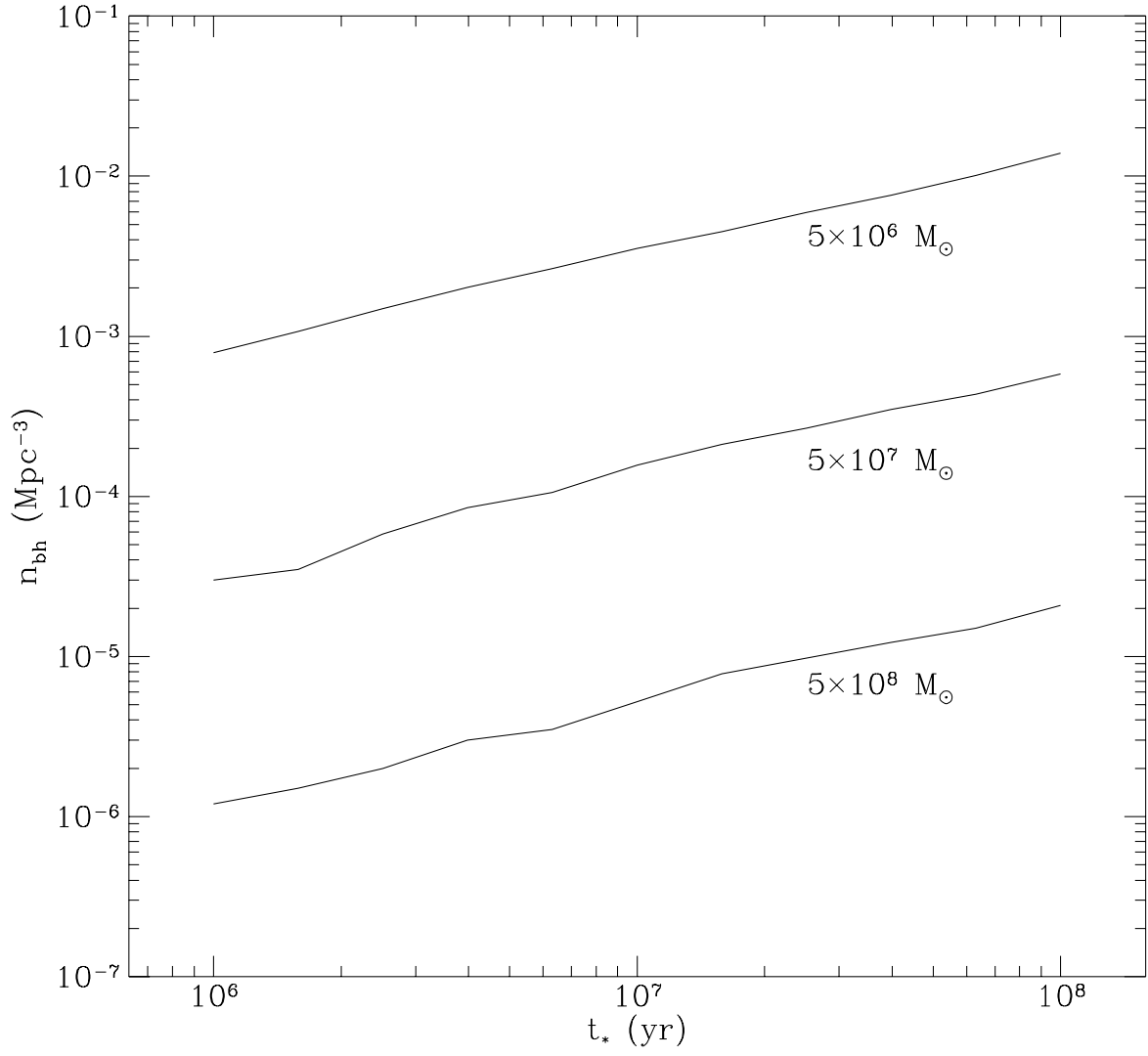

Fig. 3

# ORIGIN OF QUASAR PROGENITORS FROM THE COLLAPSE OF LOW-SPIN COSMOLOGICAL PERTURBATIONS

Daniel J. Eisenstein[1] and Abraham Loeb

Astronomy Department, Harvard University

60 Garden St., Cambridge MA 02138

## ABSTRACT

We show that seeds for quasar black holes could have originated from the initial cosmological collapse of overdense regions with unusually small rotation. The gas in these rare regions collapses into a compact disk that shrinks on a short viscous time scale. Using an analytical model, we calculate the low-spin tail of the probability distribution of angular momenta for objects that collapse out of a Gaussian random field of initial density perturbations. The population of low-spin systems is significant for any viable power spectrum of primordial density perturbations. Most objects form just above the cosmological Jeans mass $\sim 10^5 M_\odot$ at high redshifts $z \gtrsim 10$. In the standard cold dark matter cosmology, the comoving density of $10^{6-7} M_\odot$ objects with viscous evolution times shorter than $\sim 10^{6-7}$ years is $\sim 10^{-3} (h/0.5)^3$ Mpc$^{-3}$, comparable to the local density of bright galaxies. The seed black holes tend to reside within larger mass systems that collapse later and supply the gas needed for the bright quasar activity.

*Subject headings:* black hole physics–cosmology: theory–quasars: general

---

1 Also at: Physics Department, Harvard University

# 1. INTRODUCTION

The extraordinary luminosity and compact size of quasars as inferred from unresolved lensed images (Rauch & Blandford 1991; Rix et al. 1992), radio jet cores (Baath et al. 1992), and rapid x-ray variability (Remillard et al. 1991), indicate that active galactic nuclei are most probably powered by the accretion of gas onto massive black holes. Indeed, recent observations of the gas kinematics near the center of the giant elliptical galaxy M87 (Harms et al. 1994) suggest a central black hole with a mass of order $10^9 M_\odot$, as expected if this galaxy had a past quasar history. The integrated light of quasars (Sołtan 1982; Chokshi & Turner 1992) implies that a fraction $\gtrsim 3 \times 10^{-5}$ of all the baryons in the universe ended up in black holes. Formation of black holes must therefore be a non-negligible consequence of gravitational instability in the early universe, yet the origin of these $\sim 10^{6-10} M_\odot$ black holes in standard cosmologies is enigmatic (Turner 1991; Loeb 1993).

A key obstacle to forming the seeds for massive black holes in galactic centers is the centrifugal barrier. Because of its angular momentum, a typical gas cloud that collapses at high redshifts, becomes rotationally supported on a scale larger by $\sim 10^{6-8}$ than its Schwarzschild radius. While turbulent viscosity could in principle transport angular momentum outwards, thus allowing the gas distribution to contract, this process is too slow for typical galactic disks. Instead, the gas is expected to fragment and form an extended stellar system rather than a relativistic object (Loeb & Rasio 1994). In this paper, we argue that some systems acquire much less angular momenta than average during their cosmological collapse. After forming a compact gaseous disk in their initial collapse, these low-spin systems would quickly contract to form a relativistic object that serves later as a seed for a quasar black hole.

To find the fraction of systems with sufficiently low initial angular momenta, we apply a cosmological collapse model developed elsewhere (Eisenstein & Loeb 1995, hereafter Paper I). In hierarchical structure formation scenarios, collapsing objects acquire angular momentum by tidal interactions with their surroundings (Hoyle 1949; Peebles 1969). The total amount of angular momentum $J$ is often quoted in terms of a dimensionless spin parameter $\lambda = J|E|^{1/2}/GM^{5/2}$, where $E$ and $M$ are the binding energy and the mass of the object at virialization. N-body simulations (Barnes & Efstathiou 1987) find a mean value of $\langle \lambda \rangle \approx$ 0.05 that depends only weakly on the power spectrum of density fluctuations or the mass scale. Here, however, we need to study the full distribution of the angular momentum and in particular its low-spin tail. State-of-the-art N-body simulations (Warren et al. 1992) are limited by computation power from probing the rare low-spin tail of this distribution, especially on small mass scales ($\sim 10^6 M_\odot$). We therefore adopt an analytical model to calculate the angular momentum of collapsing ellipsoidal systems subject to a cosmological shear. By calculating $\lambda$ as a function of initial conditions, we then use the statistics of Gaussian random fields to compute the fraction of low-spin objects that could potentially result in quasar progenitors.

The outline of this work is as follows. In §2, we estimate the viscous contraction time of a compact gaseous disk with some initial angular momentum and mass. In §3, we describe the methods of Paper I that are used to calculate the abundance of low-spin systems as a function of their mass and angular momentum. In §4, the numerical results from this approach are combined with the results of §2 to determine the abundance of quasar seeds as a function of mass scale in a cold dark matter cosmology. Finally, §5 summarizes our main conclusions.

## 2. EVOLUTION OF COMPACT GASEOUS DISKS

In this work we consider the possibility that conditions in the initial collapse of a cosmological perturbation may favor the formation of a compact gaseous seed for a quasar black hole. We examine a collapsing region in a universe of cold dark matter particles and baryons. During the initial collapse and virialization of the system, the baryons separate from the dark matter due to their ability to shock and dissipate energy. Since the collapse is at high redshifts ($z \gtrsim 10$), the gas cools quickly via Bremsstrahlung and inverse Compton scattering off the microwave background photons. Thus, within a few dynamical times the baryons settle into a rotationally-supported thin disk. If such a disk is not disrupted and does not fragment, then turbulent viscosity transports angular momentum outwards and the disk contracts. Eventually the rate of gravitational energy being released approachs the Eddington luminosity and the disk thickens due to radiation pressure. Further contraction of the bloated disk is limited by cooling. However, at this stage the system is quite relativistic and rapidly evolves to a massive black hole (Loeb & Rasio 1994).

This sequence of events would be altered if a considerable fraction of the gas fragments into stars (Loeb & Rasio 1994). Stellar systems will not dissipate energy or transport angular momentum as efficiently, and supernovae driven winds may blow the remaining gas out of the potential well (Haehnelt & Rees 1993). The first possibility is that stars might form before the gas settles to a cold disk. However, in analogy with the over-abundance of spiral galaxies relative to ellipticals in the field, it is likely that a fair fraction of all gaseous systems collapse to a disk before they fragment into stars. The next possibility is that star formation occurs inside the disk before viscous transport of angular momentum allows the disk to shrink to a black hole. The disk will not form a quasar seed if its viscous contraction time $t_{\rm vis}$ is longer than the time for supernovae to expel the gas or the time for a dominant fraction of the gas to be converted into stars. We denote the characteristic timescale for the fastest of these processes as $t_\star$. Based on the conditions in other star formation environments $t_\star \sim 10^{6-9}$ years; therefore we will consider a broad range of values for $t_\star$ in the subsequent analysis (cf. Fig. 3). Our approach will be to estimate the viscous time for a disk of a given mass and angular momentum, and compare this time to the threshold value $t_\star$. In doing so, we will neglect the effects of non-axisymmetric modes (e.g. bars). Only disks with lower than usual angular momentum (and therefore small initial size) have a sufficiently short viscous evolution time to form a quasar seed. In §4, we will calculate the probability distribution of initial angular momenta for the compact disks considered in this section. The low-spin tail of this distribution, defined by the condition $t_{\rm vis} < t_\star$, will then be used to calculate the abundance of black hole progenitors.

For the purpose of estimating $t_{\rm vis}$ we focus on the global evolution of the disk rather than on its detailed radial structure. In addition, we tailor our analysis to the low-spin case, where the disk is small and optically thick. To begin, we consider a collapsing region with initial angular momentum $J$, total mass of dark matter and gas $M$, and gas fraction $\Omega_g$. In low-spin systems especially, the disk size is much smaller than the core radius of the dark matter. The dark matter therefore has a negligible contribution to the disk potential and we may treat the gas as self-gravitating. Under this assumption, the gas reaches rotational support at a radius $r_b \approx j^2/GM_g$, where $j \equiv J/M$ is its specific angular momentum, $G$ is Newton's constant, and $M_g \equiv \Omega_g M$ the mass of the gas. Note that here we have assumed that the gas and dark matter have the same specific angular momentum; this is the simplest case and ignores any transfer between the two components during their separation (Fall & Efstathiou 1980; see however Hernquist 1989). We then define $x_b \equiv r_b/r_{\rm Sch}$, where $r_{\rm Sch} \equiv 2GM_g/c^2$ is the Schwarzschild radius of the gas, and $c$ is the speed of light. In §3 and §4 we will derive a probability distribution of $x_b$ for an ensemble of system drawn from a random set of initial cosmological conditions (Fig. 1). For now, let us quantify the rarity of a disk size by defining $y \equiv x/\langle x_b \rangle$, where $x$ is the disk radius in units of $r_{\rm Sch}$, and $\langle x_b \rangle \equiv 1.4 \times 10^8 (M_g/5 \times 10^6 M_\odot)^{-0.6}$ is a reasonable fit to the mean value of the probability distributions in Figure 1. Typical objects have $y_b \sim 1$, while low-spin objects have $y_b \ll 1$.

The viscous timescale for the initial disk is $t_{\rm vis} \approx \rho r_b^2/\eta$, where $\rho$ is an average disk density, and $\eta$ is the viscosity coefficient. In a thin disk, the viscosity can be parametrized by a single constant $\alpha \lesssim 1$ (Shakura & Sunyaev 1973), so that

$$\eta = \alpha p \left(\frac{r^3}{GM_g}\right)^{1/2}, \tag{1}$$

where $p \approx \rho c_s^2$ is the ionized gas pressure, $c_s = (5kT/3m_p)^{1/2}$ is the sound speed, $k$ is Boltzmann's constant and $m_p$ is the proton mass. Thus, we find

$$t_{\rm vis} \approx \frac{\sqrt{2}GM_g}{\alpha c c_s^2} x_b^{1/2}. \tag{2}$$

In evaluating the viscosity (cf. Eq. (1)) we have ignored radiation pressure. If radiation pressure $p_{\rm rad}$ dominates over gas pressure, then the disk radiates energy at a rate close to the Eddington luminosity. This can be shown as follows. The radiation flux emerging from the surface of an ionized disk is $F \sim -(cm_p/\sigma_T)(\rho^{-1}[\partial p_{\rm rad}/\partial z])$, where $\sigma_T$ is the Thomson cross-section, and $\rho$ is the mass density. Hydrostatic equilibrium along the vertical $z$-axis implies $\rho^{-1}(\partial p_{\rm rad}/\partial z) \sim -2\pi G\Sigma$, where $\Sigma \sim M_g/\pi r^2$ is the surface density of the disk. By substituting this relation in the expression for $F$, we get the total luminosity of the disk, $L \sim 2\pi r^2 F \sim L_E$, where $L_E \equiv 4\pi GM_g m_p c/\sigma_T$ is the Eddington luminosity. Thus, a disk dominated by radiation pressure would radiate its gravitational binding energy, $E_{\rm grav} \sim GM_g^2/r$, on the cooling time

$$t_{\rm cool} = \frac{E_{\rm grav}}{L_E} \sim \frac{2 \times 10^8}{x} {\rm yr}. \tag{3}$$

The collapse of a primordial gas cloud with low-spin is therefore divided into two stages. Initially, the gas cloud collapses to a radius $x \gg 1$ at which $t_{\rm cool} \ll t_{\rm vis}$, and the system cools to a state where the gas pressure dominates (because viscous heating is much weaker than radiative cooling otherwise). Subsequently, the system contracts due to viscous angular momentum transport, and so the value of $x$ drops while the value of $c_s^2$ increases. Eventually, the decreasing value of $t_{\rm vis}$ approaches the increasing value of $t_{\rm cool}$, and the disk becomes dominated by radiation pressure. The later contraction of the bloated supermassive disk is limited by cooling at the Eddington rate. Because the disk is supported by radiation pressure, it resists gravitational fragmentation and star formation, and so inevitably evolves into a massive black hole (cf. Loeb & Rasio 1994). Thus, the gaseous system is vulnerable to disruption primarily during the first phase, when the potential well of the disk is relatively shallow and the Jeans mass is low due to the dominance of gas pressure. We therefore identify $t_{\rm vis}$ in equation (2) as the relevant quantity to be compared to the disruption timescale $t_\star$ (e.g., due to star formation or supernovae) in determining whether a system can become a black hole progenitor.

The only remaining unknown in equation (2) is the temperature that sets the value of $c_s^2$ in the interior of the disk. For an optically thick disk, the internal temperature is determined by the balance between the rate of gravitational energy release and the luminosity emerging from the disk surface. Assuming full ionization, the vertical optical depth of the disk to Thomson scattering is given by

$$\tau \approx \frac{\sigma_T M_g}{2\pi r^2 m_p} \approx 6 \times 10^4 \left(\frac{M_g}{5 \times 10^6 M_\odot}\right)^{0.2} \left(\frac{y}{5 \times 10^{-4}}\right)^{-2}. \quad (4)$$

If $\tau \gg 1$ for both scattering and absorption processes, the surface of the optically-thick disk radiates as a blackbody at a temperature $T_s$. However the interior temperature $T_c$ at the symmetry plane of the disk, is higher. The radiative flux at the disk surface is $F = \sigma T_s^4$, where $\sigma$ is the Stefan-Boltzmann constant. At a steady state for the vertical heat transfer, the same flux originates from the symmetry plane of the disk $F = -c(dp_{\rm rad}/d\tau) \approx \sigma T_c^4/\tau$. Equating these two flux estimates implies a central temperature $T_c \approx \tau^{1/4} T_s$. We use this central temperature to fix the sound speed in equation (2). The luminosity of the disk is ultimately provided by the gravitational binding energy release over the viscous timescale, $L \approx E_{\rm grav}/t_{\rm vis} \approx M_g c^2/2x t_{\rm vis}$, but is also given by $2\pi r^2 \sigma T_s^4$. Setting these two expressions equal closes our set of equations. Combining these equations, we get our basic result

$$t_{\rm vis} \approx 1.6 \times 10^6 \ {\rm yr} \cdot \alpha_{0.1}^{-4/3} \left(\frac{M_g}{5 \times 10^6 M_\odot}\right)^{0.6} \left(\frac{y_b}{5 \times 10^{-4}}\right)^{7/3}, \quad (5)$$

where $\alpha_{0.1} \equiv \alpha/0.1$ (cf. the limit on $\alpha$ in Narayan, Loeb & Kumar 1994). In addition,

$$T_c \approx 1.2 \times 10^4 \ {\rm K} \ \alpha_{0.1}^{1/3} \left(\frac{M_g}{5 \times 10^6 M_\odot}\right)^{0.1} \left(\frac{y}{5 \times 10^{-4}}\right)^{-11/6}. \quad (6)$$

If the gas is not fully ionized as assumed above (i.e. the temperature $\lesssim 6000$ K), then the opacity calculation is more complicated (Lenzuni, Chernoff, & Salpeter 1991). However, $t_{\rm vis} \propto \tau^{-1/3}$, so the uncertainties in the opacity of the disk due to atomic or molecular contributions have a weak influence on the estimate of its viscous time in equation (5). Note that the above equations apply only to low-spin objects since they require $\tau \gg 1$ [cf. Eq. (4)]. Typical disks with $y_b \sim 1$ are optically-thin and have viscous times greater than the age of the universe.

Without other processes taking over, the central part of a compact gaseous disk would evolve to a black hole progenitor over the viscous time given in equation (5). We should compare this timescale to the timescales for processes that might disrupt this evolution history. The disk gas may fragment into stars or be expelled by supernovae; however, these processes require a minimum time $t_\star$ to occur. We therefore require $t_{\rm vis} < t_\star$ for the formation of black holes from the gaseous disks. In §4 we will consider the consequences of different values for $t_\star$, ranging from $\sim 10^6$ yr up to the age of the universe $\sim 10^8$ yr at the typical collapse redshift $z \approx 20$. Note that disks with low viscous times are quite small and therefore are bound by deep potential wells. The depth of the potential well of the initial gaseous disk is represented by the characteristic rotational velocity $(2x_b)^{-1/2} c = 210 \,\text{km}\,\text{s}^{-1} (x_b/10^6)^{-1/2}$. Although lower mass systems have on average shallower potential wells (cf. Fig. 1), the low-spin tail for all mass scales naturally results in compact semi-relativistic disks in which the gas is better trapped against supernovae winds. For example, a $5 \times 10^6 \, M_\odot$ disk with $y \sim 10^{-3}$ has a gravitational binding energy $GM_g^2/r \sim 10^{55}$ erg, roughly equivalent to the hydrodynamic energy output of $10^4$ supernovae; i.e. it takes many simultaneous supernovae to disrupt this disk.

According to equation (5), the disk luminosity $L = E_{\rm grav}/t_{\rm vis}$ approaches the Eddington luminosity $L_E$ at a value of $y \approx 8 \times 10^{-5} \alpha_{0.1}^{0.4}$, independent of $M_g$. In this regime the system bloats into a thick disk (or equivalently, a rotating supermassive star), and further contraction is limited by the cooling time in equation (3). For initial values $y_b \gtrsim 5 \times 10^{-4}$, we find that $L \lesssim 0.002 \, \alpha_{0.1}^{4/3} \, L_E$, and a thin disk forms in the initial collapse, as assumed.

## 3. COSMOLOGICAL COLLAPSE MODEL

In §2 we considered the viscous evolution time of compact self-gravitating gaseous disks. If the disks are sufficiently compact initially, they are likely to evolve into massive black holes before being disrupted by supernovae and star formation. We next attempt to find the abundance of such systems in the universe.

The initial radius of a disk is determined by its mass and angular momentum. The system acquires its angular momentum through tidal coupling to external torques during its cosmological collapse. Most of the angular momentum is gathered when the system reaches maximum expansion and turns around, long before the baryons virialize and collapse to a disk. Different systems get different amounts of specific angular momentum, depending on the particular random realization of the density field in their environment. The abundance of compact disks is therefore related to the low-spin tail of the probability distribution of angular momenta for collapsing systems in the universe. Low-spin objects result from rare environments that produce very weak tidal forces. Since the objects in this tail are rare and compact, it is difficult to identify them in state-of-the-art numerical simulations (e.g. Warren et al. 1992) due to the limited resolution and simulated volumes. We are therefore motivated to study this low-spin tail analytically using the non-linear collapse model for cosmological perturbations described in Paper I. As shown in Paper I and demonstrated later, this model is well-suited in particular to studying the cosmological collapse of low-spin systems.

We begin at a high redshift with a Gaussian random field of linear density fluctuations that has a cold dark matter power spectrum (Bardeen et al. 1986), although our results are only weakly dependent on the shape of the power spectrum on large scales. We use $\Omega = 1$, $H_0 = 50 \, \mathrm{km \, s^{-1} \, Mpc^{-1}}$ ($h = 0.5$), and a power-spectrum normalization of $\sigma_{8h^{-1} \, \mathrm{Mpc}} = 1.0$. We consider a high density peak at the origin and divide the universe into two regions, the collapsing object and the source of the external torque acting on it. By taking a spherical boundary, the external torque spinning the object may be expanded as a multipole series. The monopole produces no torque, and the dipole corresponds to a uniform acceleration of the object and so produces no rotation about its center of mass. Hence, the quadrupole is the leading term. Previous studies found the higher multipoles to be much smaller (Ryden 1988; Quinn & Binney 1992). We use only the quadrupole tide of the background and ignore higher terms.

The evolution of the quadrupole moment of the object with time is approximated by treating the object as a homogeneous triaxial ellipsoidal overdensity above a uniform background. We choose the initial ellipsoid to be the unique homogeneous ellipsoid that matches the mass, overdensity, and quadrupole moments of the inner region. When the ellipsoid is placed in a quadratic gravitational potential, the equations of motion scale linearly with radius so that all similar shells of the ellipsoid behave uniformly (Peebles 1980). Both the ellipsoid self-gravity and the quadrupole tidal potential are of this form. The latter ingredient allows the method to include the tidal shear at the origin, the importance of which has been emphasized recently (Bertschinger & Jain 1994; Bond & Myers 1993). We then assume that the background remains smooth with a density equal to its unperturbed value through the non-linear regime (Icke 1973; White & Silk 1979). Hence the background contributes a spherical quadratic potential, which at late times is small. This approximation has been shown to be remarkably good even past turnaround in the case without external forces (White 1993). The quadrupole moments of the ellipsoid obey the correct linear regime scaling at early times, thereby exactly matching the linear quadrupole moments of the actual perturbations. In difference from previous analytical treatments (Ryden 1988; Quinn & Binney 1992), we use the non-zero initial peculiar velocity field associated with the growing mode velocities, as determined by both the ellipsoid and the tidal shear.

Under the above approximations, the dynamics of the ellipsoid is reduced to a set of nine second-order ordinary differential equations, which can be easily integrated given a random set of initial conditions and the time dependence of the tidal potential (see paper I for details). As a consequence of virialization, we artificially prevent any further contraction of an axis beyond 40% of its maximum length (Bond & Myers 1993). The precise value of this threshold has little effect on the quadrupole moments. We stop evolving the system when the time is equal to the collapse time of a spherical object with the same initial overdensity. Since the tidal field has already been included in the dynamics, the ellipsoid rotates and one simply measures its angular momentum at the end. The vorticity of the object remains zero before the artificial virialization, since all forces are derived from a potential.

Previous analytical work has assumed that because the object is a high density peak, the background tidal torques can be described by linear perturbation theory. We find this assumption unjustified, as the region just outside the boundary of the object has a similar overdensity to that of the object. We therefore evolve the background torque beyond the linear regime by dividing the exterior region into twenty spherical shells centered at the origin. The initial radii of the shells are a fixed set of multiples of the radius of the boundary between the object and the background. These radii evolve with time according to the mass interior to them (Gunn & Gott 1972). Each shell has a mean initial overdensity $\overline{\delta}_0$ and carries a separate contribution to the tidal torque. The overdensity of each shell $\delta(z)$ is computed as a function of redshift $z$ and the tidal force exerted on the ellipsoid is scaled by $(1+z)\delta(z)/(1+z_0)\overline{\delta}_0$ above linear theory. This scaling matches linear theory at early times but provides more torque at late times, when the torquing material is closer to the ellipsoid than the Hubble expansion would imply.

The initial conditions are the mean overdensity and quadrupole moments of the inner region and the surrounding twenty shells. The joint probability distribution for these may be computed in an exact form, given a Gaussian random field of primordial density perturbations (cf. Appendix A in Paper I). We then place certain restrictions on the allowed range of initial conditions to make them more suitable to our approximations. The inner region is required to have an overdensity of at least $2.5\sigma$, where $\sigma = (\delta M/M)_{rms}$ for this mass scale. This makes it more likely that the object is the first in its neighborhood to collapse. To favor formation of bound objects, the initial shear is constrained not to cause any part of the ellipsoid to have a peculiar velocity that is radially outward. We also require that the initial overdensity of the tidal shells is less than 95% of the ellipsoid overdensity, so that these shells do not collapse before the end time. For $10^8 M_\odot$ regions, only 0.40% of all regions satisfy all of the above restrictions, primarily because overdensities above $2.5\sigma$ occur only 0.62% of the time. Higher mass scales have slightly larger acceptance rates.

In agreement with N-body simulations, our model yields $\langle \lambda \rangle \approx 0.04$ for all mass scales between $10^7$ and $10^{12}$ $M_\odot$, independent of the normalization of the power spectrum. When the integration ends, typical objects have undergone collapse along two axes, but their long axis is still near turnaround. This is likely to result in orbit crossing with the collapsing background shells. *However, this complicated situation is avoided in low-spin objects.* Because the shear is small, these objects are more spherical during their evolution, and their long axis collapses in a time close to the spherical top-hat prediction (cf. Fig. 8 in Paper I). These systems collapse before the tidal shells and thereby avoid the complex shell-crossing phenomena that affect typical objects. Our approximate treatment of the quadrupole torque is therefore well-suited to studying the low-spin tail of the probability distributions of angular momenta. There is, however, a background of objects that at the final time are very elongated and yet by chance have a low angular momentum. These objects should not be considered black hole candidates, as the gas is settling into a long filament rather than a compact disk. Figure 2 shows the distribution of axis ratio (long over short) at the end time for all objects with angular momentum lower than some threshold. The threshold for this sample was fixed so that $t_{\rm vis} < 10^7$ years in equation (5). One can see that more than half of all low-spin objects have axis ratios less than two, making them good candidates for becoming black hole progenitors. For simplicity, we will include all objects with sufficiently low angular momenta in the results presented in the next section, regardless of their final shape.

## 4. RESULTS AND DISCUSSION

According to equation (5), the viscous time of systems with $y_b \lesssim 10^{-3}$ is comparable to a star formation time or to the time it takes a supernova to form out of the first generation of stars. More important, $t_{\rm vis}$ depends on $y_b$ to a high power. Therefore, the critical value of $y_b$ for the formation of a black hole progenitor is only weakly dependent on the uncertain values of $t_\star$ or $\alpha$. Because $t_{\rm vis} \propto M_g^{0.6}$, we expect more candidates for black holes at lower masses. The lower bound on $M_g$ is the cosmological Jeans mass of $\sim 10^5 M_\odot$, at which the effects of gas pressure and thermal feedback strongly influence the dynamics of the collapse (Peebles & Dicke 1968; Peebles 1993). For each mass scale, a threshold value for $t_{\rm vis}$ defines a maximum $y_b$ for black hole formation; we then find the fraction of systems with $y_b$ lower than this value. For example, the condition that the viscous time of a system with $M_g = 5 \times 10^6$ M$_\odot$ be smaller than $10^7$ years is that $y_b$ is less than $1.1 \times 10^{-3}$. This corresponds to a disk of size $\sim 10^{17}$ cm, smaller than average by about $10^3$, with a rotational velocity of 550 km s$^{-1}$, a surface density of order $10^9$ M$_\odot$ pc$^{-2}$, and a specific angular momentum smaller than average by a factor of 30.

To calculate the number densities of black hole progenitors, we begin by noting that the number density of regions of mass $M$ is simply $n_0 = \rho_c/M$. These regions represent all possible initial conditions; however, as described in the previous section, we only work with a small fraction $A \approx 0.40\%$ of these, since we require high-$\sigma$ peaks. We conservatively assume that all the initial conditions that we reject do not lead to black hole progenitors. After evolving many examples of the accepted initial conditions, we denote the fraction with sufficiently low angular momentum (low $y_b$) as $f$. The number density of low-spin objects is then $n_0 A f$. For our results, we evolve $2 \times 10^5$ realizations of accepted initial conditions for each of several total (baryonic + dark matter) mass scales: $10^8$ M$_\odot$, $10^9$ M$_\odot$, $10^{10}$ M$_\odot$, and $10^{12}$ M$_\odot$. In all cases, we use a viscosity parameter $\alpha = 0.1$, a gas mass fraction $\Omega_g = 0.05$, and a bias parameter of unity. The probability distributions for $x_b$ are shown in Figure 1. The comoving densities $n_{\rm bh}$ of black hole progenitors with different masses are plotted in Figure 3, as a function of the viscous time threshold $t_\star$ in the constraint $t_{\rm vis} < t_\star$. Regardless of stellar processes, the value of $t_\star$ should not exceed the time scale for cosmological infall of additional mass with higher specific angular momentum. This upper bound corresponds to the Hubble time $\sim 10^8$ yr at the typical collapse redshift $z \approx 20$. At the low mass end, the number densities are comparable to the comoving density of bright galaxies (Peebles 1993), $n_\star = 1.25 \times 10^{-3 \pm 0.2} (h/0.5)^3$ Mpc$^{-3}$. The masses of the resulting black holes are assumed to be $M_g \equiv \Omega_g M$; this is an overestimate because a fraction of the gas must carry off the angular momentum of the initial disk to infinity. The curves in Figure 3 are very nearly power laws with $n_{bh} \propto t_\star^{0.62}$.

The number densities in Figure 3 have a variety of dependences on the parameters of the model. First, because $t_{\rm vis} \propto \Omega_g^{-4.07}$, a reduction in the fraction of gas that ends in the disk (e.g. due to star formation during the initial infall or a different value of the baryon density parameter $\Omega_b$) would increase $t_{\rm vis}$ considerably. Second, the number densities in Figure 3 use only a small subset ($\geq 2.5\sigma$) of all the available regions for each mass scale; lowering the peak threshold would increase the number of black holes. However, the ellipsoid model is more applicable for isolated peaks, which suggests high-$\sigma$ peaks. In addition, there is an anti-correlation of peak height and angular momentum, so that requiring high-$\sigma$ peaks makes for an efficient search. Third, since the candidates listed must have angular momenta lower by at least an order of magnitude than the mean, it is possible that subleading effects in the calculation of $\lambda$ are important at this level. While this may change some of the numbers in Figure 3, the existence of a significant low-spin tail is in qualitative agreement with the broad probability distributions obtained in N-body simulations (Warren et al. 1992). Fourth, $\langle x_b \rangle$ is inversely proportional to the mean collapse redshift, which in turn depends on the unknown amplitude of the power spectrum on small mass scales. Higher collapse redshifts (i.e. lower bias) result in denser objects and shorter viscous times.

In the ellipsoid model, we focus on a collapsing region as a complete object, but real systems collapse over a range of time, with continuing accretion forming larger and larger objects. In this ongoing process, the angular momentum of the object is continually changing in time. How correct is it, then, to use the angular momentum of a fixed mass scale (equivalent to freezing the collapse in time) as the input to our model for the evolution of the gaseous disk? First, the contraction of the thin disk occurs rapidly, with a time scale much smaller than the time scale for accretion to significantly alter its angular momentum. Second, once a system with low angular momentum has undergone free fall to a small radius, it will have a very high surface density and small cross-section compared to any clumps of new accreting material (of order $\sim 10^6$ times higher), and thus be relatively unaffected by collisions with such clumps. It therefore seems reasonable that a compact disk will contract without interference from the new infall.

In this model, we have neglected the role played by non-axisymmetric perturbations such as bars. In fact, because the low-spin gaseous disks studied here are self-gravitating, one expects the systems to be unstable to bar formation as they cool to low temperatures. These instabilities act on the dynamical timescale

$$t_{\rm dyn} \approx \sqrt{\frac{3\pi}{G\rho}} \approx 2 \times 10^2 \text{ yr} \left(\frac{y}{5 \times 10^{-4}}\right)^{3/2} \left(\frac{M_g}{5 \times 10^6 \, M_\odot}\right)^{0.1}, \qquad (7)$$

which is far shorter than the viscous timescale. While bars do generally transport angular momentum outward and allow material to move inward (Larson 1994), it is not clear that this can lead directly to the formation of massive black holes (see, for example, the simulations by Loeb & Rasio 1994, where bar formation did not result in a massive gaseous core near the center). Rather, we expect the bar to heat and thicken the disk, possibly fragmenting it into clumps. Unlike typical galactic conditions, the time scale in equation (7) is too short for stars to form and turn the disk into a collisionless system. With pressure ocassionally stabilizing the newly-heated disk, turbulent viscosity is again the mechanism responsible for further contraction. A detailed calculation of the viscous transport of angular momentum would depend on the complicated structure resulting from the bar, and so we rely the simplified $\alpha$-viscosity model presented in §2 to estimate the viscous time.

Within our model it is not possible to calculate the initial mass function of quasar black holes. Although two regions of the same mass are correctly sampled from the underlying density field, one cannot study different mass scales at the same time. There is no guarantee that two low-spin objects of slightly different masses are not actually the same perturbation with slightly different boundaries. While this overcounting might be small when considering very different mass scales, the predicted number density of black holes drops quickly with mass. Thus, most of the seed population of black holes would form just above the Jeans mass at relatively high redshifts ($z \approx 20$). This result is consistent with the lowest values of the empirically determined black hole masses in active galactic nuclei (Peterson 1993; Padovani, Burg, & Edelson 1990). Moreover, Loeb & Rasio (1994) have argued recently that quasar black holes can grow by stable accretion in galactic bulges only if their seed mass is $\gtrsim 10^{5-6} M_\odot$, and therefore postulated the existence of a primordial population of massive seeds. The coincidence between the Jeans mass and this lower limit allows our model to provide just this postulated population.

## 5. CONCLUSIONS

The appearance of quasar black holes at high redshifts is naturally linked to a seed population of compact gaseous systems, more massive than the cosmological Jeans mass of $\sim 10^5 M_\odot$. While compact gas clouds of this type are not common due to the centrifugal barrier of typical systems in the universe (Loeb & Rasio 1994), they can still result from the initial collapse of rare systems with unusually low angular momentum. Such systems originate in environments where the surrounding cosmological density field induces an unusually low shear during their collapse. The gas in these low-spin systems settles into a compact semi-relativistic disk, whose viscous time is sufficiently short to avoid disruption by either star formation or supernovae.

In this work we have calculated the abundance of low-spin systems in the universe, assuming a Gaussian random field of initial density perturbations in a standard cold dark matter cosmology. The resulting number density of progenitors for primordial $\sim 10^{6-7} M_\odot$ black holes is estimated to be somewhat larger or comparable to the density of bright galaxies (cf. Fig. 3). After their cosmological collapse, these compact progenitors have initial disk radii less than $10^{17}$ cm and rotational velocities greater than $500\,\mathrm{km\,s^{-1}}$, leading to viscous evolution times shorter than $10^{6-7}$ yr. If the resulting black holes were to sink to the center of galaxies by dynamical friction, they would later serve as seeds for quasar activity. In particular, their initial mass $\gtrsim 10^6 M_\odot$ would be sufficiently high to dominate gravity near the center and to stabilize a steady accretion flow from the gas reservoirs of their host galactic bulges (Loeb & Rasio 1994).

Because of the strong dependence of the viscous time of the initial disk on its specific angular momentum, most gaseous disks in the universe fragment into stars long before their angular momentum is transported away by viscosity. For typical systems, the resulting stellar relaxation time is much longer than the Hubble time at the relevant redshifts. However, low-spin objects that for some reason were unable to avoid star formation, would still result in compact stellar systems. These dense systems have short relaxation times and could dynamically evolve to additional black holes (Quinlan & Shapiro 1990).

If a seed black hole of mass $\gtrsim 10^6 M_\odot$ is initially located inside a galactic bulge, dynamical friction will bring it to the center on a time shorter than the age of the universe (Binney & Tremaine 1987). On the other hand, if the black hole is located far away from the nucleus (e.g. in the galactic halo) it will not spiral inwards and the low accretion rate of gas onto it will be unobservable. As shown in Appendix A of Paper I, a high-$\sigma$ peak on the $M_g = 10^{6-7}\,\mathrm{M_\odot}$ mass scale is very likely to be surrounded by a high density peak on the galactic bulge mass scale. For example, we find that a $3\sigma$ peak on the $10^8 M_\odot$ (total mass) scale constrains the density field to have a $2.2\sigma$ peak on average for the $10^{10}$ $M_\odot$ spherical region centered on the first peak. Thus, the seed black holes that form out of high-$\sigma$ peaks are very likely to reside inside galactic bulge systems and therefore sink to their center of mass by dynamical friction. The observed peak in the quasar population at $z \approx 2$ may simply reflect the epoch of galaxy formation when considerable infall feeds these seed black holes (Haehnelt & Rees 1993). The decline in the quasar density at lower redshifts would then result from the dilution of their gas supply.

The observed mass function of quasar black holes should be very different from its initial state due to accretion. Since a typical quasar luminosity $L_q$ corresponds to an accretion rate of 1.7 $M_\odot \,\mathrm{yr}^{-1}$ $(L_q/10^{46} \mathrm{\ ergs\ s}^{-1})(\epsilon/0.1)^{-1}$, where $\epsilon$ is the conversion efficiency of accreted mass into radiation, it is possible to get considerable variations in the final black hole masses due to differences in the fuel reservoir of their host galaxies. These differences are generic, as the seed black holes considered in this work are likely to be surrounded by galactic mass systems that collapse later and feed them with additional gas, thus resulting in the bright quasar activity.

We thank Eyal Maoz, Eve Ostriker, and Jerry Ostriker for useful discussions. D.J.E. was supported in part by a National Science Foundation Graduate Research Fellowship.

# FIGURE CAPTION

**Fig. 1**: Probability distribution $P(x_b)$ for the centrifugal barrier radius of the gas in units of the Schwarzschild radius, $x_b \equiv r_b/r_{\text{Sch}}$. The three histograms correspond to different total mass systems. The gas fraction is assumed to be 5% of the total masses listed.

**Fig. 2**: Probability distribution for the ratio of long axis length to short axis length at the final time. The broad histogram (dashed line) includes the full sample of objects with a total (dark matter and baryonic) mass of $10^8$ M$_\odot$ The sharp histogram (solid line) includes only the low-spin objects defined by the condition $t_{\text{vis}} < t_\star = 10^7$ yrs.

**Fig. 3**: Number density of black hole progenitors for three different mass scales as a function of the viscous time threshold $t_\star$ in the condition $t_{\text{vis}} < t_\star$. The densities are for a standard cold dark matter cosmology with $\Omega = 1$, $h = 0.5$.